\documentclass[12pt,twocolumn,tighten]{aastex63}
\turnoffedit

\usepackage{apjfonts}
\usepackage{url}
\usepackage{hyperref}
\usepackage{natbib}
\usepackage{amsmath,amstext,amssymb}
\usepackage[caption=false]{subfig} % for subfloat
\usepackage{bm} % for bold hat accents
\usepackage{xcolor, fontawesome}
\usepackage{color}

\newcommand{\minus}{\scalebox{0.5}[1.0]{$-$}}

\newcommand{\eg}{e.g.,}

\newcommand{\gitlink}{\href{https://github.com/lgbouma/gyro-interp}{github.com/lgbouma/gyro-interp}}

\newcommand{\caltech}{Department of Astronomy, MC 249-17, California
Institute of Technology, Pasadena, CA 91125, USA}

\begin{document}

\title{The Empirical Limits of Gyrochronology}

\correspondingauthor{Luke G. Bouma}
\email{luke@astro.caltech.edu}

\received{2023 Jan 25}
\revised{2023 Mar 3}
\accepted{2023 Mar 14}
\shorttitle{Gyrochronology by Interpolation} 

\shortauthors{Bouma, Palumbo, Hillenbrand}

\author[0000-0002-0514-5538]{Luke~G.~Bouma}
\altaffiliation{51 Pegasi b Fellow}
\affiliation{\caltech}

\author[0000-0001-7967-1795]{Elsa~K.~Palumbo}
\affiliation{\caltech}

\author{Lynne~A.~Hillenbrand}
\affiliation{\caltech}

\begin{abstract}
  The promise of gyrochronology is that given a star's rotation period
  and mass, its age can be inferred.  The reality of gyrochronology is
  complicated by effects other than ordinary magnetized braking that
  alter stellar rotation periods.  In this work, we present an
  interpolation-based gyrochronology framework that reproduces the
  time- and mass-dependent spin-down rates implied by the latest open
  cluster data, while also matching the rate at which the dispersion
  in initial stellar rotation periods decreases as stars age.  We
  validate our technique for stars with temperatures of 3800--6200\,K
  and ages of 0.08--2.6\,gigayears (Gyr), and use it to reexamine the
  empirical limits of gyrochronology.  In line with previous work, we
  find that the uncertainty floor varies strongly with both stellar
  mass and age.  For Sun-like stars ($\approx$5800\,K), the
  statistical age uncertainties improve monotonically from $\pm$38\%
  at 0.2\,Gyr to $\pm12$\% at 2\,Gyr, and are caused by the empirical
  scatter of the cluster rotation sequences combined with the rate of
  stellar spin-down.  For low-mass K-dwarfs ($\approx$4200\,K), the
  posteriors are highly asymmetric due to stalled spin-down, and
  $\pm$1$\sigma$ age uncertainties vary non-monotonically between 10\%
  and 50\% over the first few gigayears.  High-mass K-dwarfs (5000\,K)
  older than $\approx$1.5\,Gyr yield the most precise ages, with
  limiting uncertainties currently set by possible changes in the
  spin-down rate (12\% systematic), the calibration of the absolute
  age scale (8\% systematic), and the width of the slow sequence (4\%
  statistical).  An open-source implementation, \texttt{gyro-interp},
  is available online at \gitlink.
\end{abstract}


\section{Introduction}
\label{sec:intro}

The ages of stars are fundamental for our understanding of planetary,
stellar, and galactic evolution.  Unfortunately, stellar ages are not
directly measurable, and so the astronomical age scale is tied to a
mix of semifundamental, model-dependent, and empirical techniques
\citep{Soderblom_2010}.  One empirical age-dating method is to use a
star's spin-down as a clock \citep{Kawaler_1989,Barnes_2003}.  This
gyrochronal technique leverages direct measurements of stellar surface
rotation periods, typically inferred from photometric modulation
induced by spots or faculae.  The clock's mechanism is magnetized
braking that drives rotation periods to increase as the square root of
time \citep{Weber_1967, Skumanich_1972}.  While data from open
clusters have shown the limitations of this approximation, the idea
has been useful, and it has set the foundation for many empirical
studies of how rotation period, age, and activity are interrelated
\citep[\eg][]{Noyes_1984,Barnes_2007,Mamajek_2008,Barnes_2010,Angus_2019,Spada_2020}.

This work aims to clarify the accuracy and precision of gyrochronology
for stars on the main-sequence.  Our main impetus for writing was the
realization that available models did not match observations of open
cluster rotation periods
\citep[\eg][]{Curtis_2019_ngc6811,Curtis_2020}.  The disagreement was
most severe for K-dwarfs, which have stellar rotation rates that stall
from 0.7 to 1.4 Gyr \citep{Agueros_2018,Curtis_2020}.  While a likely
physical explanation centers on the timescale for angular momentum
exchange between the radiative core and convective envelope
\citep{Spada_2020}, accuracy is paramount because any bias in the
rotation models propagates into bias on the inferred ages.

Regarding precision, previous analytic studies have reported age
uncertainties for field FGK dwarfs of 13--20\% \citep{Barnes_2007},
and have noted that these uncertainties increase for young stars due
to larger empirical scatter in their rotation sequences
\citep{Barnes_2010}.  The question of how this empirical scatter,
often described as ``fast'' and ``slow'' sequences in the
rotation--color plane, limits gyrochronal precision was analyzed in
detail by \citet{Epstein_2014}.  For stars older than 0.5\,Gyr, their
approach was to consider the range of possible ages that a star with
fixed rotation period and mass might have, and to convert this range
into an age uncertainty.  Our work formalizes this idea.  If an
astronomer wishes to infer the age of an individual field star, they
do not know whether their star is on the fast or slow sequence. They
simply know the star's rotation period and mass, and so they must
marginalize over the population-level scatter in order to determine a
posterior probability distribution for the age.  Ultimately,
\citet{Epstein_2014} emphasized that this type of approach needed
empirical guidance in order to mitigate the systematic uncertainties
in the spin-down models; such guidance now exists.

Using the latest available open cluster data
(Section~\ref{sec:clusters}), we calibrate a new gyrochronal model
that interpolates between the open cluster rotation sequences
(Section~\ref{sec:model}).  Given a star's rotation period, effective
temperature, and their uncertainties, our framework returns the
implied gyrochronal age posterior, which is often asymmetric
(Section~\ref{sec:results}).  We validate our model against both
training and test data, and focus our discussion and conclusions
(Section~\ref{sec:conclusions}) on the empirical limits of gyrochronal
age-dating.  An open-source implementation is available online at
\gitlink.

\begin{deluxetable*}{lccccccc}[!t]
\tabletypesize{\small}
\tablecaption{Reference clusters and parameters used for the core gyrochrone calibration. \label{tab:clusters}}
\tablenum{1}

\tablehead{
  \colhead{Name} & \colhead{Reference Age} & \colhead{Age Provenance} & \colhead{$A_V$} & \colhead{$A_V$ Provenance} & \colhead{Instrument} & \colhead{$P_{\rm rot}$ Provenance} & \colhead{Recovered Age$^\star$}
}

\startdata
\hline
  $\alpha$ Per 	 & $79.0^{+1.5}_{-2.3}$\,Myr    & (1)  & 0.28  & (2$^\dagger$) & TESS    &  (2)    & $70^{+8}_{-8}$\,Myr   \\
  Pleiades 		   & $127.4^{+6.3}_{-10.0}$\,Myr  & (1)  & 0.12  & (3)           & K2      &  (4)    & $117^{+6}_{-6}$\,Myr    \\
  Blanco-1 		   & $137.1^{+7.0}_{-33.0}$\,Myr  & (1)  & 0.031 & (5)           & NGTS    &  (5)    & $134^{+11}_{-9}$\,Myr   \\
  Psc-Eri stream & Pleiades-coeval         			& (6)  & 0     & (6)           & TESS    &  (6)    & $137^{+11}_{-10}$\,Myr    \\
  NGC-3532 			 & $300\pm50$\,Myr 						  & (7)  & 0.034 & (8)           & Y4KCam  &  (8)    & $306^{+11}_{-9}$\,Myr  \\
  Group-X 			 & $300\pm60$\,Myr 							& (9)  & 0.016 & (9)           & TESS 	 &  (9)    & $260^{+29}_{-27}$\,Myr    \\
  Praesepe 			 & $670 \pm 67$\,Myr 					  & (10) & 0.035 & (3)           & K2 		 &  (11)   & $738^{+22}_{-20}$\,Myr  \\
  NGC-6811 			 & $1040 \pm 70$\,Myr 			    & (12) & 0.15  & (3)           & K2 		 &  (12)   & $946^{+16}_{-15}$\,Myr  \\
  NGC-6819 			 & $2.5 \pm 0.2$\,Gyr 					& (13) & 0.44  & (3)           & Kepler  &  (14)   & $2518^{+32}_{-33}$\,Myr \\
  Ruprecht-147 	 & $2.7 \pm 0.2$\,Gyr 			 		& (15) & 0.30  & (3)           & K2      &  (3) 	 & $2637^{+51}_{-52}$\,Myr \\
\enddata
\tablecomments{
References:
(1) \citet{GalindoGuil_2022};
(2) \citet{Boyle_2022};
(3) \citet{Curtis_2020};
(4) \citet{Rebull_2016};
(5) \citet{Gillen_2020};
(6) \citet{curtis_2019};
(7) \citet{Fritzewski_2019};
(8) \citet{Fritzewski_2021};
(9) \citet{Messina_2022};
(10) \citet{Douglas_2019};
(11) \citet{Rampalli_2021};
(12) \citet{Curtis_2019_ngc6811};
(13) \citet{Jeffries_2013};
(14) \citet{Meibom_2015};
(15) \citet{Torres_2020}.
%\citet{GalindoGuil_2022} ages are from the lithium depletion boundary (LDB);
%ages for the older clusters come from either isochrone-fitting
%(NGC-3532), differential gyrochronology (Group-X, Praesepe, NGC-6811), or
%detached eclipsing binaries (NGC-6819, Ruprecht-147).
$^\dagger$The adopted $\alpha$ Per reddening varies across the
cluster, per \citet{Boyle_2022}; this table reports
the median value.
$^\star$See Section~\ref{sec:validation}.
}
\vspace{-0.7cm}
\end{deluxetable*}

\section{Benchmark Clusters}
\label{sec:clusters}

\subsection{Rotation Data}
\label{subsec:clusterdata}

To calibrate our model, we first collected rotation period data from
open clusters that have been surveyed using precise space and
ground-based photometers.  The clusters that we examined are listed in
Table~\ref{tab:clusters}, along with their ages and $V$-band
extinctions.  These clusters were selected based on the completeness
of available rotation period catalogs for F, G, K, and early M dwarfs.
The Pleiades, Blanco-1, and Psc-Eri were concatenated as a
120\,megayear (Myr) sequence, since their rotation--temperature
sequences were visually indistinguishable.  The upper age anchor,
Ruprecht-147, was similarly combined with NGC-6819 to make a 2.6\,Gyr
sequence.  While older populations have been studied
\citep{Barnes_2016, Dungee_2022}, their rotation--color sequences do
not yet have sufficient coverage to be usable in our core analysis.
Our lower anchor, $\alpha$~Per, was selected based on its converged
rotation--temperature sequence above 0.8\,$M_\odot$
\citep{Boyle_2022}.  Our model is therefore only constrained between
80\,Myr and 2.6\,Gyr.

\subsection{Effective Temperatures}
\label{subsec:teff}

For our effective temperature scale, we adopted the
\citet{Curtis_2020} conversion from dereddened Gaia Data Release 2
(DR2) $G_{\rm BP} - G_{\rm RP}$ colors to effective temperatures.
This calibration was determined using FGK stars with high-resolution
spectra \citep{Brewer_2016}, nearby stars with interferometric radii
\citep{Boyajian_2012}, and M-dwarfs with optical and near-infrared
spectroscopy \citep{Mann_2015}.  The typical precision in temperature
from this relationship is 50\,K for stars near the zero-age
main-sequence (ZAMS).  We explicitly used Gaia DR2 mean photometry to
calculate the temperatures, since the intrinsic difference between the
Gaia DR2 and DR3 colors is important at this scale.  For all other
Gaia-based quantities in our analysis, we used the DR3 values.  For
the extinction corrections, we adopted the reddening values listed in
Table~\ref{tab:clusters}.  We dereddened the observed Gaia DR2 $G_{\rm
BP} - G_{\rm RP}$ colors by assuming $E(G_{\rm BP} - G_{\rm RP}) =
0.415 A_{\rm V}$, similar to \citet{Curtis_2020}.  

\subsection{Binarity Filters}
\label{subsec:binarityfilters}

Binarity can affect the locations of stars in rotation--color space by
observationally biasing photometric color measurements, and also by
physically altering stellar rotation rates through \eg\ tidal spin-up
or early disk dispersal.  To remove possible binaries from our
calibration sample, we applied the following filters to each cluster
dataset.

{\it Photometric binarity}---We plotted the Gaia DR3 color--absolute
magnitude diagrams in $M_{\rm G}$ vs. $G_{\rm BP} - G_{\rm RP}$, $G -
G_{\rm RP}$, and $G_{\rm BP} - G$, and manually drew loci to remove
over or under-luminous stars in each diagram. 
% We called this flag \texttt{flag\_camd\_outlier}.

{\it RUWE}---We examined diagrams of the Gaia DR3 renormalized
unit weight error (RUWE) as a function of brightness, and based on
these diagrams required ${\rm RUWE}>1.2$.  Outliers in this space can
be caused by astrometric binarity, or by marginally resolved
point-sources fitted with a single-source PSF model by the
Gaia pipeline.  %\texttt{flag\_ruwe\_outlier} was set if ${\rm
%RUWE}>1.2$.

{\it Radial velocity scatter}---We examined diagrams of Gaia DR3
``radial velocity error'' as a function of $G$-mag.  Since this
quantity is the standard deviation of the Gaia RV time-series,
outliers can imply single-lined spectroscopic binarity.  We manually
removed such stars. %; they are noted
%with the \texttt{flag\_rverror\_outlier} flag.

{\it Crowding}---We queried Gaia DR3 to determine how many stars were
within 1 instrument pixel distance of each target star (\eg\ 4$''$/px
for Kepler).  Any stars within $\approx$20$\times$ the brightness of
the target star ($\Delta G < 3.25$) were noted, and the target stars
were removed from further consideration.  Although not all visual
companions are binaries, their presence can complicate rotation period
measurements, particularly in cluster environments.

{\it Gaia DR3 Non-Single-Stars}---Gaia DR3 includes a column to flag
known or suspected eclipsing, astrometric, and spectroscopic binaries.
We directly merged against this column to remove such sources.

{\it Final calibration sample}---The combination of the filters
described above yields the set of stars that show no evidence for
binarity or crowding.  However, some of the rotation period analyses
in Table~\ref{tab:clusters} include additional relevant quality flags.
For instance, light curves showing multiple photometric periods can
indicate unresolved binarity.  We used all relevant filters available
from the original authors if they were designed to select single stars
with reliable rotation periods.  The final combination of these
filters with our own flag for possible binarity yields our sample of
benchmark rotators. %, which are highlighted using the
%\texttt{flag\_benchmark\_period} flag.

\begin{figure*}[t]
	\begin{center}
		\leavevmode
		\subfloat{
			\includegraphics[width=0.60\textwidth]{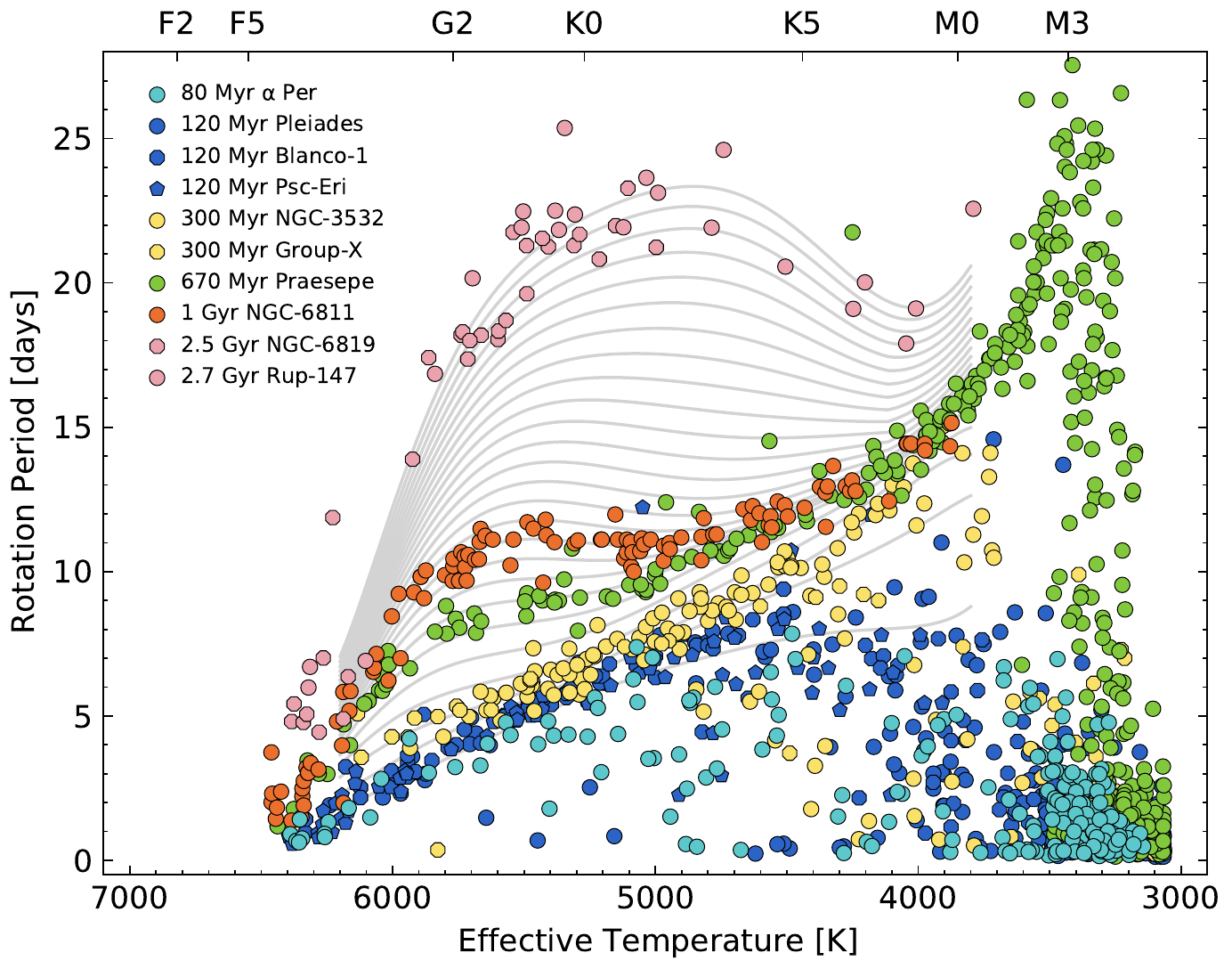}
		}
		
		\vspace{-0.45cm}
		\subfloat{
			\includegraphics[width=0.60\textwidth]{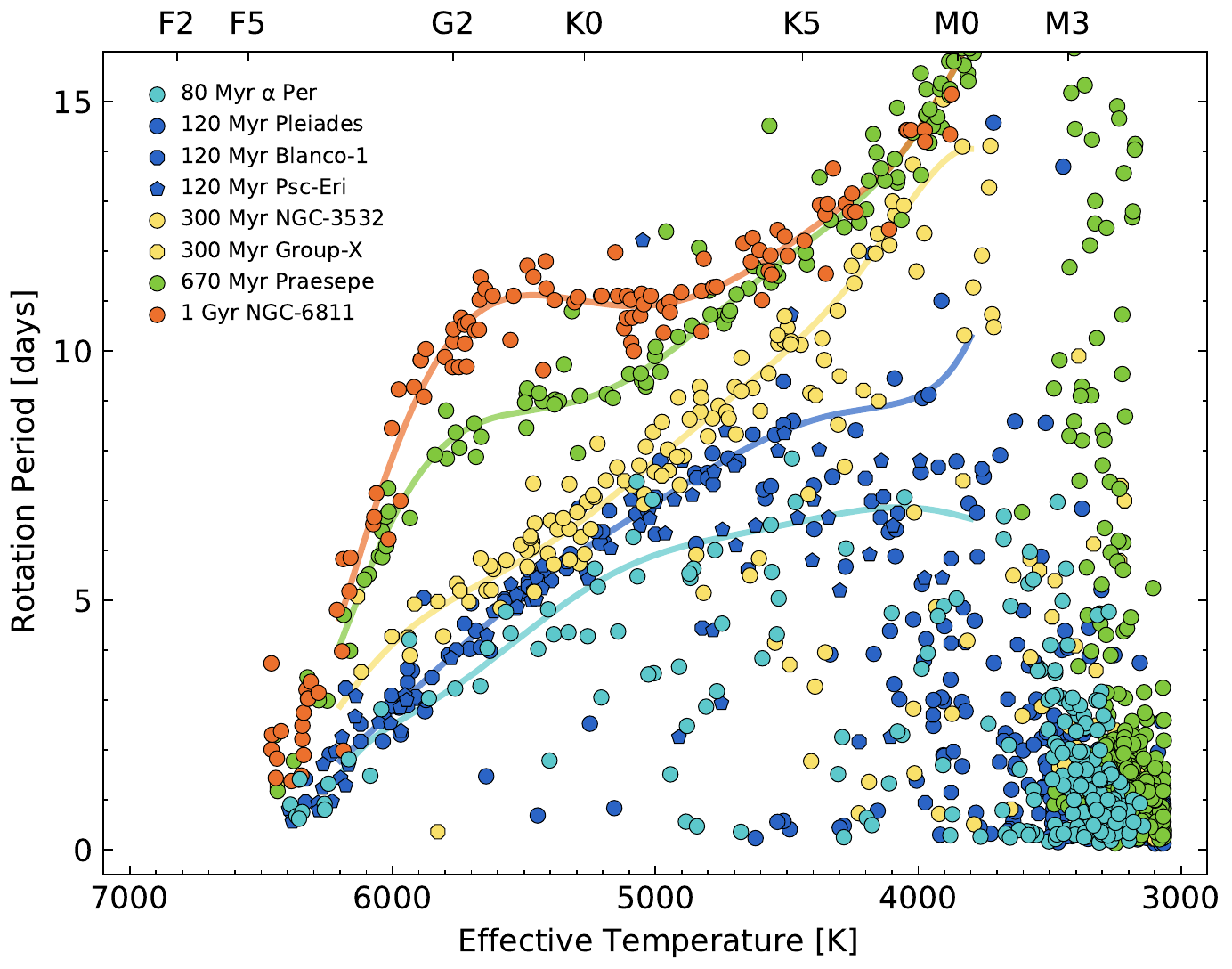}
		}
	\end{center}
	\vspace{-0.55cm}
	\caption{
    {\bf Open cluster data and models.}
    The top panel shows the data that we aim to model, and the bottom
    panel focuses on the first gigayear.  Gray lines in the top panel
    show the mean model for the rotation period distribution, and are uniformly spaced at integer multiples
    of 100\,Myr.  They are evaluated using a seventh-order polynomial
    for each cluster (colored lines, bottom panel), and interpolated
    piecewise between those reference loci.  The model is defined over
    temperatures of 3800--6200\,K, and ages of 0.08--2.6\,Gyr.  Data
    behind the Figure are available as a machine-readable table.
    \label{fig:clusters}
	}
\end{figure*}

\subsection{The Single-Star Calibration Sequence}

Figure~\ref{fig:clusters} is the result of the data curation process
described in Sections~\ref{subsec:clusterdata}
through~\ref{subsec:binarityfilters}.  While we have omitted the
possible binaries described in Section~\ref{subsec:binarityfilters}
for visual clarity, they are included in the Data behind the Figure.
The gray lines are derived from polynomial fits that we describe in
the following section.  Comparing against the rotation--color
sequences in say \citet{GodoyRivera_2021}, it is impressive how sparse
the fast sequence is for hot stars.  In the 120\,Myr clusters, both
Blanco-1 and Psc-Eri have no apparently single fast rotators hotter
than 5000\,K.  The Pleiades has four.  The rapid rotator sequence is
similarly sparse at 300\,Myr.  The large binary fraction of
fast-sequence stars warrants future analysis, to understand whether
the binary separations and mass ratios for these systems are typical
of the field binary population.

\section{A Gyrochronology Model}
\label{sec:model}

Here we present a model that aims to accurately describe the evolving
rotation period distributions of F7--M0 dwarfs with ages of
0.08--2.6\,Gyr.  The goal is to then use this model to assess the
precision with which rotation periods can be used to infer ages.  To
perform this analysis, our model needs to account for the trends
visible in Figure~\ref{fig:clusters}: stellar spin-down rates vary
with both mass and age; stellar spin-down can stall; and higher-mass
stars younger than Praesepe tend to converge to the slow sequence
before lower-mass stars.  Our approach will ultimately use
interpolation, based on the logic that there are certain regions of
Figure~\ref{fig:clusters} in which a hypothetical star located between
two cluster sequences would need to have an age intermediate to those
two clusters.  A few formalities are needed to make this idea
rigorous.

\subsection{Formalism}
For a given star, we have an observed rotation period
$\bm\tilde{P}_{\rm rot}$ and stellar effective temperature
$\bm\tilde{T}_{\rm eff}$ with measurement uncertainties
$\sigma_{\bm\tilde{P}_{\rm rot}}$ and $\sigma_{\bm\tilde{T}_{\rm
eff}}$. Given these data, we want to find the posterior probability
distribution for the age $t$ of the star. We write the corresponding
probability density as $p(t|\bm D)$, where $\bm D = \{ \tilde{P}_{\rm rot}, \bm\tilde{T}_{\rm
	eff} \}$ are the observed data.   We find $p(t|D)$ by marginalizing over the joint
probability density $p(t, P_{\rm rot}, T_{\rm eff} | D)$, where $P_{\rm rot}$ and  $T_{\rm eff}$ are the true
rotation period and temperature of the star.  Mathematically, this means
\authorcomment1{(NOTE DURING EDIT1: Here and for the remaining manuscript, the probability
	notation was updated from the mathematician's $f_{x|y}$ to the astronomer's $p(x|y)$.  We also omitted the implicit uncertainties ``$s$'', and wrote all instances of the observables as $D$.}
\begin{align}	
p(t| D)	
=	
\iint 	
p(P_{\rm rot}, T_{\rm eff}, t | D)	
\ {\rm d}P_{\rm rot}\ {\rm d}T_{\rm eff}.	
\label{eq:theintegral}	
\end{align}	
By Bayes' rule, the integrand can be written as	
\begin{align}	
p(P_{\rm rot}, T_{\rm eff}, t | D)	
& \propto 	
p(D | P_{\rm rot}, T_{\rm eff}, t)
\cdot	p( P_{\rm rot}, T_{\rm eff}, t )	
\end{align}	
where the first term is the likelihood and the latter is the prior.	

\subsection{Likelihood}
\label{subsec:likelihood}

For the likelihood, we assume that the observed rotation period and
temperature have Gaussian uncertainties and are measured
independently.  In this case,
\begin{align}
    p(D | P_{\rm rot}, T_{\rm eff}, t )
    &= p(\bm\tilde{P}_{\rm rot} | P_{\rm rot}, T_{\rm eff}, t ) 
    \cdot
    p(\bm\tilde{T}_{\rm eff} | P_{\rm rot}, T_{\rm eff}, t ),
\end{align}
and the temperature and rotation period distributions are specified by
$\bm\tilde{T}_{\rm eff} \sim \mathcal{N}(T_{\rm eff}, \sigma_{\bm\tilde{T}_{\rm eff}}^2)$ and 
$\bm\tilde{P}_{\rm rot} \sim \mathcal{N}(P_{\rm rot}, \sigma_{\bm\tilde{P}_{\rm rot}}^2)$,
where $\mathcal{N}$ denotes the normal distribution.  In other words,
our likelihood is a product of two normal distributions.

\subsection{Prior}
\label{subsec:prior}

The prior is more interesting.  By the chain rule, 
\begin{align}
    p( P_{\rm rot}, T_{\rm eff}, t )
    &=
    p(P_{\rm rot} | T_{\rm eff}, t) \cdot \ p(T_{\rm eff}) \cdot\ p(t),
    \label{eq:prior}
\end{align}
where we have assumed $p(T_{\rm eff}|t) = p(T_{\rm eff})$ because in
our model, changes in stellar temperature through time are ignored.
We assume that age and temperature are uniformly distributed,
$ t \sim \mathcal{U}(t_{\rm min}, t_{\rm max})$
and
$T_{\rm eff} \sim \mathcal{U}(T_{\rm eff}^{\rm min}, T_{\rm eff}^{\rm max})$,
where $(t_{\rm min}, t_{\rm max})$, $(T_{\rm eff}^{\rm min}, T_{\rm
eff}^{\rm max})$ are the limiting ages and temperatures for our model, respectively.
We adopt limiting ages of 0 to 2.6\,Gyr, and limiting temperatures of
3800 to 6200\,K.  The upper limit on age is set by the oldest clusters
in our dataset (Table~\ref{tab:clusters}), and the temperature limits
are set to include the regions in which stellar rotation is most
correlated with age.  While one might imagine a prior on temperature
informed by the stellar initial mass function, or a prior on age
informed by the star formation history of the Milky Way, the star
formation rate has been approximately constant over the past 10~Gyr \cite[\eg][]{2004A&A...418..989N} and incorporating a
stellar mass function would systematically bias already accurate
measurements towards lower temperatures.  We do not consider such
additions. 	

The remaining term in Equation~\ref{eq:prior}, $p(P_{\rm rot} |
T_{\rm eff}, t)$, is the core of our model.  We propose a functional
form for $p(P_{\rm rot} | T_{\rm eff}, t)$ that relies on two
components.  The first component, $\mu_{\rm slow}(T_{\rm eff}, t)$, is
the rotation period of the star if it were exactly on the slow
sequence --- this is colloquially the ``mean'' gyrochronal model for a
star's rotation period prescribed at any age and temperature.  The
second component is the {\it residual} to that mean model --- the
probability distribution for how far the star's rotation period is
from the slow sequence at any given age and temperature.  This model
parametrization is motivated by how the observed abundance of rapid
rotators changes as a function of both stellar temperature and age.

\paragraph{The Mean Model}

To parametrize the slow sequence, we fitted rotation periods in each
reference cluster with an $N^{\rm th}$ order polynomial over
3800--6200\,K.   We manually selected the slow sequence stars to
perform this fit using the data behind Figure~\ref{fig:clusters}.  We
investigated the choice of $N$ between 2 and 9, and settled on $N=7$
as a compromise between overfitting and accurately capturing the
structure of the Praesepe and NGC-6811 sequences.
While lower-order polynomials provide acceptable fits for the
80--300\,Myr clusters \citep[\eg][Appendix~A]{Curtis_2020}, for
purposes of homogeneity across all clusters we adopted a single
polynomial order. 

To model the evolution of the slow sequence, we considered a few
possible approaches, all based on interpolating between the fitted
polynomials (see Appendix~\ref{app:interp}).  We ultimately chose at
any given temperature to fit 1-D monotonic cubic splines in rotation
period as a function of age.  This guarantees a smooth increase in the
slow sequence envelope while also fitting all available data.
Systematic uncertainties associated with this choice are described in
Section~\ref{subsec:systematicunc}.  This procedure yielded the gray
lines in Figure~\ref{fig:clusters}.  At times below 80\,Myr,
we do not extrapolate; we instead let the ``mean model'' $\mu_{\rm
slow}(t, T_{\rm eff})$ equal the lowest reference polynomial rotation
period values as set by $\alpha$~Per.  This yields posterior
distributions that are uniformly distributed at ages below 80\,Myr.  Possible
options regarding extrapolation for older stars are discussed in
Appendix~\ref{app:interp}.

\begin{figure*}[!t]
	\begin{center}
		\leavevmode
		\includegraphics[width=.96\textwidth]{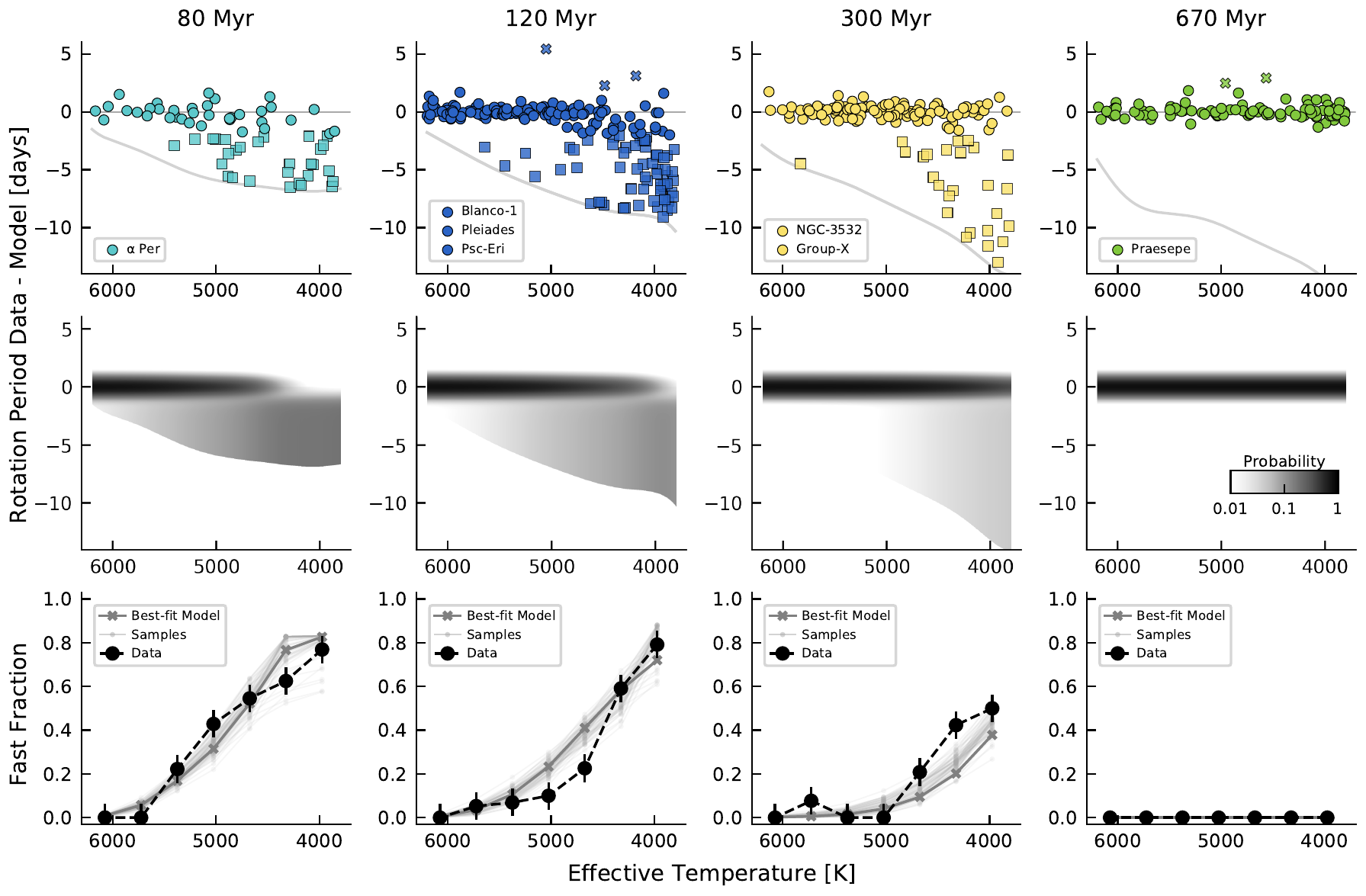}
	\end{center}
	\vspace{-0.6cm}
	\caption{
		{\bf Data, model, and goodness-of-fit.} 
    {\it Top}: Cluster rotation periods, minus the corresponding
    slow-sequence mean model at each cluster's age.  The lower gray
    envelope corresponds to a zero-day rotation period.  {\it Middle}:
    Probabilistic model for rotation period as a function of age and temperature
    (Equation~\ref{eq:residual}), fitted to the 120\,Myr, 300\,Myr,
    and 670\,Myr clusters.  
    {\it Bottom}: Fraction of stars in 350\,K bins that rotate
    ``fast'', as a function of temperature.  ``Fast'' and ``slow''
    stars are squares and circles on the top panel;  ``very slow''
    outliers are the crosses.  ``Slow'' stars show a uniform scatter
    of $\sigma$$\approx$0.51\,days around the mean model at $t\geq
    120$\,Myr.  The assumed uncertainties for Praesepe are smaller
    than the markers (see Section~\ref{subsec:fitting}).
		\label{fig:residual}
	}
\end{figure*}
\paragraph{The Residual}

The top row of Figure~\ref{fig:residual} shows the residuals for the
calibration clusters with $t\leq 670$\,Myr, relative to the polynomial
model.  Our ansatz is to model this distribution as a sum of a
Gaussian and a uniform distribution, with each distribution smoothed
around a time-dependent transition location in effective temperature.
This procedure ignores the few positive outliers.  

Mathematically, this means that the rotation period, given the age and
temperature, is drawn from
\begin{align}
  \label{eq:residual}
  P_{\rm rot}
  \sim
  &\ a_0\, \mathcal{N}_{P}(\mu_{\rm slow}, \sigma^2)
  \otimes L_{T}(T_{\rm eff}^{\rm cut}(t), k_0)
  \\
    \nonumber
  &+ a_1 g(t)\ \mathcal{U}_{P}(0, \mu_{\rm slow})
  \otimes \left( 1 - L_{T}(T_{\rm eff}^{\rm cut}(t), k_1) \right),
\end{align}
where $\mathcal{N}$ is a normal distribution, $\mathcal{U}$ is a
uniform distribution, $a_0$ and $a_1$ are scaling constants, and
$L(\ell, k)$ is the logistic function specified by a location $\ell$
and smoothing scale $k$.  Visual examples are given in the middle row
of Figure~\ref{fig:residual}.  The subscripts, for instance
$\mathcal{N}_P$, indicate the dimension over which the distribution is
defined --- period ($P$) or effective temperature ($T$), and $\otimes$
denotes an outer product.  We have also hidden the dependence of
$\mu_{\rm slow}$ on time and temperature for simplicity of notation.

The first term in Equation~\ref{eq:residual} parametrizes the slow
sequence using a Gaussian centered on $\mu_{\rm slow}(t,T_{\rm eff})$,
with a universal width $\sigma = 0.51\,{\rm days}$ set by the
observations of clusters at least as old as the Pleiades.  The
location parameter of the logistic function, $T_{\rm eff}^{\rm
cut}(t)$, is a function that monotonically decreases to account for
the age-dependent transition between the slow and fast sequence.
While other functional
forms are possible, we assumed that at any given time this function is
defined as the temperature of the lowest-mass star that has just
arrived at the main-sequence, since this is the time at which the
star's surface rotation rate is no longer affected by gravitational
contraction.  We evaluated this quantity through linear interpolation
over the solar-metallicity MIST grids \citep{Choi_2016}.  At 80, 120,
and 300\,Myr this yielded $T_{\rm eff}^{\rm cut}$ values of 4620,
4150, and 3440\,K, respectively.

\subsection{Free Parameters}
The free parameters in the model are as follows. In the residual term,
there are the amplitudes $(a_0, a_1)$, the two scaling parameters
$(k_0, k_1)$, and the slope of the linear amplitude decrease $g(t)$
for the ``fast sequence'' term through time. This would yield five
free parameters, but $a_0$ and $a_1$ are degenerate, so there are
really only four degrees of freedom.  We fixed the other terms in the
model that could in principle be allowed to vary. These included the
polynomial terms in the slow sequence model $\mu_{\rm slow}$, the
scatter around the slow sequence  $\sigma$, and the function
specifying the decrease of the effective temperature cutoff through
time $T_{\rm eff}^{\rm cut}(t)$.

\subsection{Fitting the Model}
\label{subsec:fitting}

To compare the model (Equations~\ref{eq:theintegral}
through~\ref{eq:residual}) to the data, we performed the following
procedure.  For the reference clusters at 120\,Myr ($N_\star = 196$),
300\,Myr ($N_\star = 133$), and 670\,Myr ($N_\star = 100$), we divided
the data into seven bins, starting at 3800\,K, with uniform bin widths
of 350\,K.  Including $\alpha$~Per (80\,Myr; $N_\star = 65$) as an
optional fourth dataset yielded similar results, so we omitted it for
simplicity.  In each bin, we counted the number of stars on the slow
sequence, and the number of stars on the fast sequence. We considered
a star to be ``slow'' if it is within two days of the mean slow
sequence model, and ``fast'' if it is more than two days faster than
the same model.  This cutoff was determined based on the uniform
scatter of $\sigma\approx0.51\,$days seen around the slow-sequence for
clusters with $t\geq 120$\,Myr.  We then use the resulting counts to
define a ``fast fraction,'' $F$, the ratio of fast rotating stars to
the total number of stars observed in any given temperature bin. 

The bottom row of Figure~\ref{fig:residual} shows this fast fraction
as a function of temperature.  We calculated the same summary
statistic for our model through numerical integration. This yields a
$\chi^2$ metric,
$
  \chi^2 = \sum_i  (F_i - F_{i,{\rm model}})^2 / \sigma_i^2,
$
where the sum $i$ is over the three reference sets of open clusters.
For the $\sigma_i$, the default Poissonian uncertainties would
disfavor the small number of stars from 4500--6200\,K in Praesepe that
are all on the slow sequence.  Since auxiliary clusters with similar
ages such as the Hyades \citep{Douglas_2019} and NGC-6811
\citep{Curtis_2019_ngc6811} also have fully converged slow sequences,
we adopted a prescription for the $\sigma_i$ in which we set them to
be equal to one another at 120 and 300\,Myr and ten times smaller at
670\,Myr.  This forces the model to converge to the fast sequence by
the age of Praesepe.  The normalization of the uncertainties was then
allowed to float in order to yield a reduced $\chi^2$ of unity.

We fitted the model by sampling the posterior probability using
\texttt{emcee} \citep{Foreman-Mackey13}.  We sampled over five
parameters: $a_1/a_0$, $\ln k_0$, $\ln k_1$, the slope of $g(t)$, and
the multiplicative uncertainty normalization.  The function $g(t)$ was
set to unity below 120\,Myr, and to decrease linearly to zero while
intersecting 300\,Myr at a particular value, $y_{\rm g}$.  The latter
value was the free parameter used to fit the slope of the line.  The
maximum-likelihood values yielded by this exercise were $\{a_1/a_0,
\ln k_0, \ln k_1, y_{\rm g}\} = \{8.26, \minus 4.88, \minus 6.24,
0.67\}$.  To evaluate the posterior, we assumed a prior on each
parameter that was uniformly distributed over a wide
boundary.\footnote{$a_1/a_0 \sim \mathcal{U}(1, 20)$, $\ln k_0 \sim \mathcal{U}(-10, 0)$, $\ln k_1 \sim \mathcal{U}(-10, 0)$, $y_{\rm g} \sim \mathcal{U}(0.1, 1)$.}  We
checked convergence by running the chains out to a factor of 300 times
longer than the autocorrelation time.  The resulting median parameters
and their 1$\sigma$ intervals were
$\{a_1/a_0, \ln k_0, \ln k_1, y_{\rm g}\} = \{
9.29_{\minus 2.41}^{+3.62},
\minus 4.27_{\minus 1.52}^{+2.56},
\minus 6.15_{\minus 0.25}^{+0.23},
0.63_{\minus 0.07}^{+0.03}
\}$.
The lower row of Figure~\ref{fig:residual} shows the best-fit model
plotted over 64 samples. Qualitatively, the model fits the fast	
fraction's behavior well in both temperature and time.  To check
the accuracy of our uncertainties, we performed a cross-validation
analysis in which we randomly dropped 20\% of the stars in the reference clusters,
without replacement, and then refitted the data.  The resulting parameters
all fell within the stated 1$\sigma$ uncertainty intervals.

\subsection{Evaluating the Posterior}
\label{subsec:posterior}

For any given star, we numerically evaluate
Equation~\ref{eq:theintegral} using the composite trapezoidal rule.
For each age in a requested grid, we define linear grids in the
dimensions of temperature and $y\equiv P - \mu_{\rm slow}$, each with
side length $N_{\rm grid}$.  The integration is then performed over
${\rm d}T_{\rm eff}$ and ${\rm d}y$ at each specified age.  Runtime
scales as $\mathcal{O}(N_{\rm grid})$, and takes under a minute on a
typical laptop.  This runtime estimate however assumes that the four
hyperparameters, $a_1/a_0, \ln k_0, \ln k_1,$ and $y_{\rm g}$, are
fixed.  Since these parameters are unknown, the most rigorous approach
for age inference for any one star requires sampling from the
posterior probability distribution for the hyperparameters.  Each
sample then yields its own posterior for the age from
Equation~\ref{eq:theintegral}, from which sub-samples can be drawn.
All the sub-samples can then be combined to numerically yield an
average posterior.

The top panels of Figure~\ref{fig:precision} show the results of this
sampling procedure in dotted lines, plotted underneath an alternative:
simply adopting the best-fit model (solid lines).  The results are
similar, although there are differences for most rapidly rotating
stars.  While the sampling procedure is relatively simple to
parallelize, it is a factor of $\approx$10$^3$ times more expensive
than using the best-fit model; for most practitioners, the rigor is
unlikely to justify the runtime cost.  As we will discuss in
Section~\ref{subsec:systematicunc}, this model has other systematic
uncertainties that are more important.

\section{Results}\label{sec:results}

\subsection{Model Validation}\label{sec:validation}

\authorcomment1{NOTE DURING EDIT1: This paragraph has been reworded with an updated procedure.
	Quoted ``recovered age'' values in
		the final column of Table 1 have also been updated using the hierarchical bayesian
		framework.}
As a validation test, we calculated gyrochronal age
posteriors for all 3800--6200\,K stars in Figure~\ref{fig:clusters}.
To infer the implied age for each cluster, we ``stack'' the
posteriors using \texttt{PosteriorStacker}\footnote{See
\url{github.com/JohannesBuchner/PosteriorStacker}, and Appendix~A of
\citet{Baronchelli_2020}.  This approach is viable because
\texttt{gyro-interp} adopts a uniform prior over age, and so the
hierarchical likelihood simplifies to a product of the likelihoods for
each star.}, which considers two hierarchical Bayesian models for the
intrinsic age distribution of each cluster: a Gaussian, and a
non-parametric histogram.  After omitting a few extreme
outliers\footnote{ TIC~44647574 in Psc-Eri; EPIC~212008710 in
Praesepe; KIC~5026583 and KIC~5024122 in NGC-6819; and EPIC~219774323
in Ruprecht-147}, the two approaches give similar results, and so in
the ``recovered age'' column of Table~\ref{tab:clusters} we report the
median and uncertainty of the mean cluster age assuming that the
individual stellar ages in each cluster are drawn from a Gaussian.
The resulting ages agree with the literature ages for every cluster to
within 2$\sigma$, as we would expect for a sample of ten clusters.

As an additional test, we repeated the exercise, but using data for
two open clusters outside of our training data: M34
($\approx$240\,Myr; \citealt{Meibom_2011}) and M37 ($\approx$500\,Myr;
\citealt{Hartman_2009}).  For M34, fitting the data after applying the
binarity filters described in Section~\ref{subsec:binarityfilters}
yielded an age of $222\pm20$\,Myr.  For M37, the same procedure
yielded $463\pm18$\,Myr.  The latter estimate agrees with the
isochronal age found by \citet{Hartman_2008} without convective
overshoot ($485 \pm 28$\,Myr), and is 2.5$\sigma$ below their
isochronal age that included convective overshoot ($550 \pm 30$\,Myr).

\subsection{Precision of Gyrochronology}

\begin{figure*}[!t]
	\begin{center}
		\leavevmode
		\subfloat{
			\includegraphics[width=0.31\textwidth]{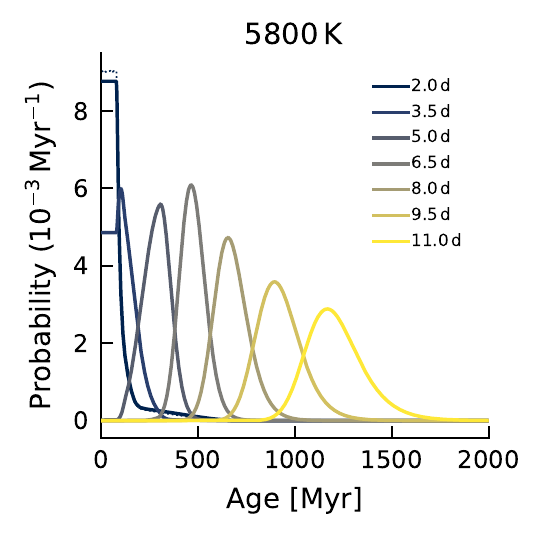}
			\includegraphics[width=0.31\textwidth]{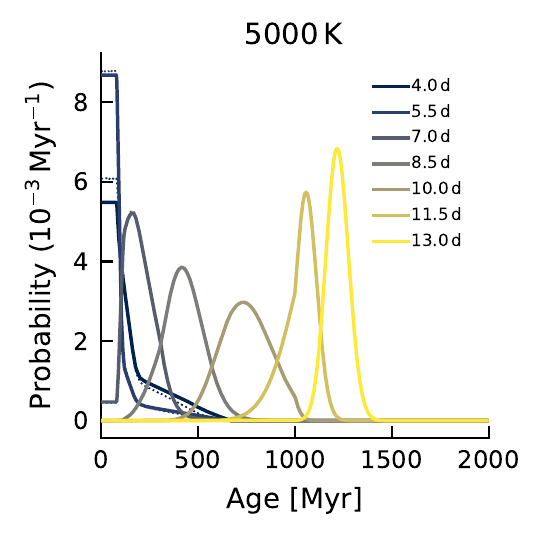}
			\includegraphics[width=0.31\textwidth]{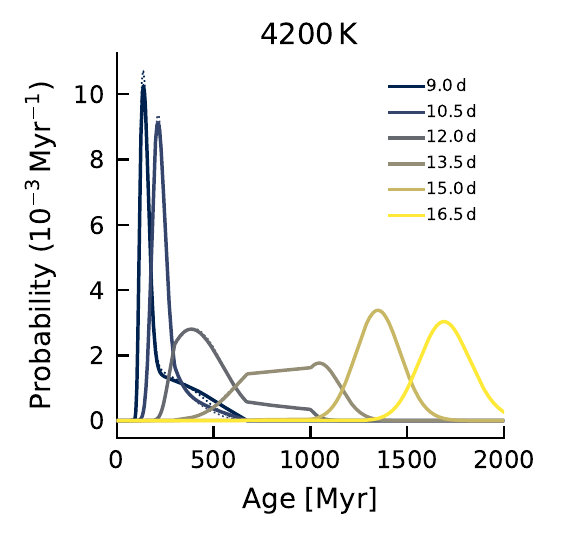}
		}
		
		\vspace{-0.35cm}
		\subfloat{
			\includegraphics[width=0.91\textwidth]{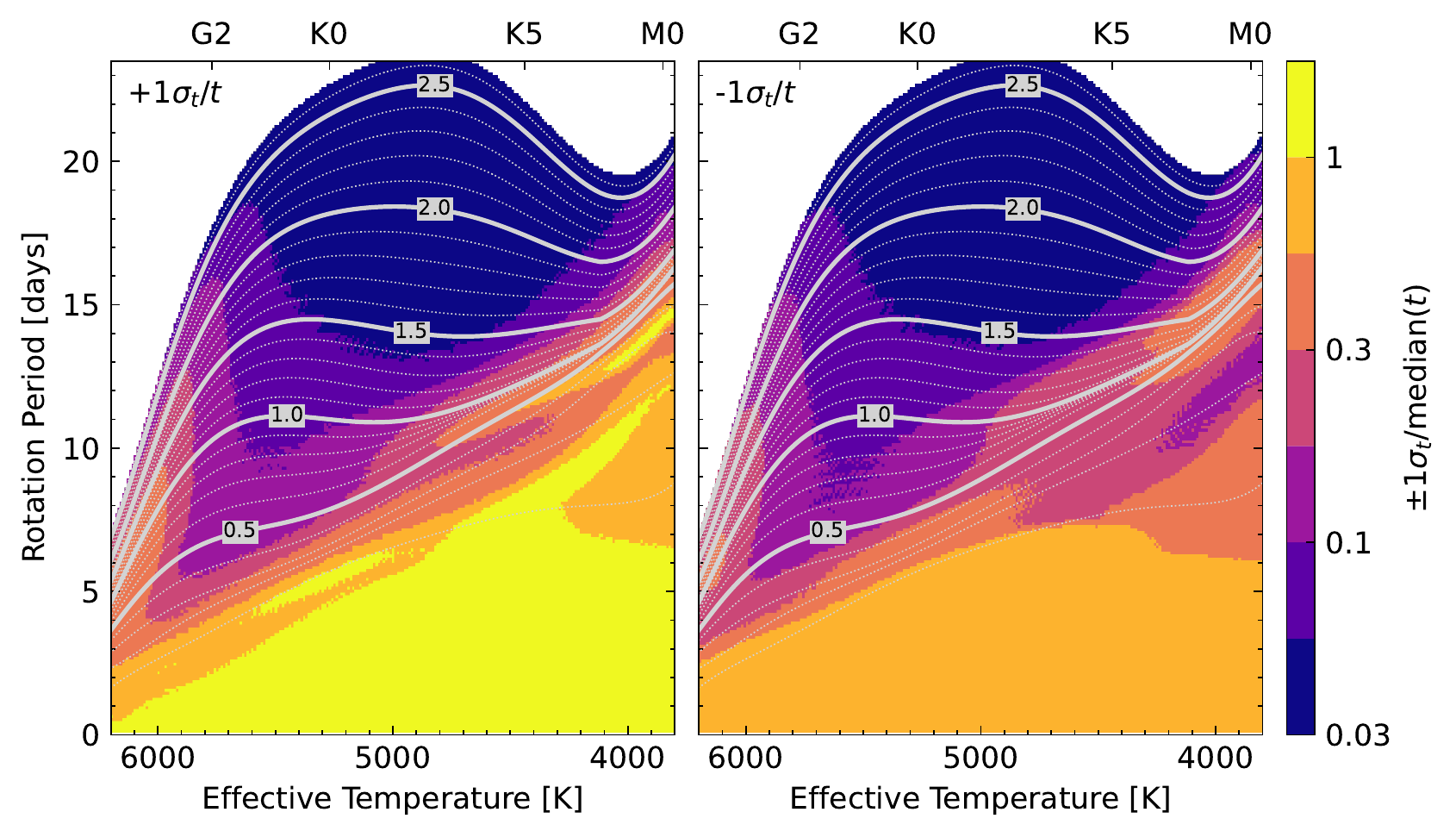}
		}
	\end{center}
	\vspace{-0.6cm}
	\caption{
		{\bf Precision of gyrochronal ages from our method. } 
    {\it Top:} Age posteriors across rotation--temperature space. In
    each subplot, each line represents a pair of $P_{\rm rot}$ and
    $T_{\rm eff}$, and assumes a precision of 50\,K in effective
    temperature, and 1\% in rotation period.  The solid lines come
    from the best-fit model in Figure~\ref{fig:residual}.  The
    under-plotted dotted lines come from a more rigorous approach that
    samples over the population-level hyperparameters discussed in
    Section~\ref{subsec:posterior}.
    {\it Bottom:} $+$1$\sigma$ ({\it left}) and $-$1$\sigma$
    uncertainty ({\it right}) of the age posterior, normalized by the
    median value.  For instance ``$\pm 1\sigma_t / {\rm median}(t) =
    0.3$'' corresponds to 30\% relative precision. Thick gray lines
    are at integer multiples of 500\,Myr, and dotted lines are spaced
    every 100\,Myr; labels are in units of gigayears.
		\label{fig:precision}
	}
\end{figure*}

Having demonstrated that our method can recover the ages of known
cluster stars, here we examine its statistical limits for individual
field stars.  The bottom panel of Figure~\ref{fig:precision} shows the
$\pm 1$$\sigma$ uncertainties, normalized by the median of the
gyrochronal age posteriors, over a grid of rotation periods and
temperatures.  Broadly speaking, the regions in which rotation periods evolve the least, such as the hottest stars and stalled $\approx$1\,Gyr
K-dwarfs, have the worst inferred precisions.

The top panel of Figure~\ref{fig:precision} visualizes vertical slices
of the bottom panel for a few canonical cases.  For a Sun-like star
($\approx$5800\,K) in its early life, the rotation period is only
informative in that it provides an upper limit on the star's age.  As
the star ages, the age posterior becomes two-sided, with a best-case
statistical precision of $\pm12\%$ at 2\,Gyr.  For a low-mass K-dwarf,
the evolution of the age posterior is more complicated.  These stars
only converge to the slow sequence by the age of Praesepe.  Their
spin-down is then observed to stall, which leads to highly asymmetric
posteriors between ages of 0.5--1.3\,Gyr.  For instance, a 4000\,K
star on the slow-sequence at 200\,Myr has a $+1\sigma$ uncertainty of
88\%, and a $\minus 1\sigma$ uncertainty of 13\%.  Nonetheless,
statistical age precisions for such stars are predicted to improve
after the era of stalled spin-down, reaching $\pm$9\% by 2\,Gyr.  The
implication is that the rotation periods of such stars can be
predictive of age, but only at certain times.

\section{Discussion \& Conclusions}
\label{sec:conclusions}

\subsection{The Gyrochronal Precision Floor}
\label{subsec:success}

A key simplifying factor in our analysis is that we assumed the
scatter of rotation periods around the slow sequence, $\sigma\approx
0.51\,{\rm days}$, is fixed in time.  Based on the data, $\sigma$
appears to be constant between 120\,Myr and 1\,Gyr
(Figure~\ref{fig:residual}, top panel).  In $\alpha$~Per, the scatter
is larger (0.85\,days), likely because the stars are only just
converging to the slow sequence.  The scatter is also larger in the
Ruprecht-147 data, but this is likely due to observational uncertainty
in the period measurements.  This empirical $\approx$0.51-day
scatter could come from a number of sources, including differential
rotation on stellar surfaces \citep{Epstein_2014}, uncertainties in
the effective temperature scale, or differing wind strengths between
stars of the same mass and age.

Regardless of the scatter's origin, it sets the floor for gyrochronal
precision, in tandem with the intrinsic spin-down rates.  In line with
previous results \citep{Barnes_2007}, gyrochronal ages for F-dwarfs
are less precise than for G-dwarfs, because F-dwarfs spin down more
slowly.  However in detail, Figure~\ref{fig:precision} shows that such
statements depend on both mass and age.  More broadly,
Figure~\ref{fig:precision} also implies that accounting for the
evolving dispersion of the rotation period distributions is a required
ingredient for producing accurate age uncertainties.

\subsection{Systematic Uncertainties}
\label{subsec:systematicunc}

The uncertainties described thus far have been statistical, rather
than systematic.  Key systematic uncertainties include the
time-varying nature of the spin-down rate, the accuracy of the
absolute age scale, and stellar binarity.

Regarding the spin-down rate, our interpolation approach guarantees
accuracy near any given reference cluster.  However, far from the
reference clusters, the choice of interpolation method can affect the
inferred ages.  We estimated the associated systematic uncertainties
by evaluating grids of $P_{\rm rot}$ vs.\ $T_{\rm eff}$ analogous to
Figure~\ref{fig:precision}, but assuming {\it i)} piecewise linear
interpolation, and {\it ii)} piecewise cubic hermite interpolating
polynomials calibrated only on the 0.08--2.6\,Gyr data (see
Appendix~\ref{app:interp}).  The difference in the medians of the age
posteriors relative to our default interpolation method is an
indicator of the systematic uncertainty.  This procedure showed a
$<$1\% bias in the inferred ages for 5000--6200\,K stars younger than
1\,Gyr, due to the dense sampling of the calibration clusters.  For
cooler stars however, a linear spin-down rate would yield differences
of up to $\pm$15\%  in the median age due to the rapid spin-down from
0.1--0.3\,Gyr (see Figure~\ref{fig:clusters}).  For older stars
between 1 and 2.6\,Gyr, the cubic interpolation yielded $\pm$100\,Myr
differences, while the linear interpolation yielded $-$200 to
$+$50\,Myr differences, with the largest differences again for stars
cooler than 4500\,K.  The summary is that from 1--2.6\,Gyr, there is a
6--12\% systematic uncertainty, with the maximum uncertainty at
1.8\,Gyr, half-way between the two reference clusters.

Regarding the absolute age scale, Table~\ref{tab:clusters} reports age
precisions for the calibration clusters of 3--20\%, with the largest
uncertainties for the $300\pm60$\,Myr NGC-3532 and Group-X.  To assess
how shifts in this scale might affect our gyrochronal ages, we again
calculated grids of $P_{\rm rot}$ vs.\ $T_{\rm eff}$, but in this case
we shifted all the reference cluster ages by either $+$1$\sigma$ or
$-$1$\sigma$.  The results showed what one would naively expect: if
all the clusters are $\pm$$1$$\sigma$ older than their reference ages,
then the changes in the inferred ages match however
much freedom there is in the local age scale.  For example, for a
$5800$\,K star with a 5.1\,day rotation period, our method
statistically yields $t = 308^{+70}_{-81}$\,Myr, roughly the age of
NGC-3532.  However, the age of that cluster is uncertain at the 20\%
level, and so the median age from our estimate for this worst-case
scenario could be systematically shifted either up or down by
$\pm$20\% to match the true age of the reference cluster.  From the
uncertainties quoted in Table~\ref{tab:clusters}, and from comparable
studies in the literature \citep[\eg][]{Dahm_2015}, the age scale
itself seems to currently be defined at a $\sim$10\% level of
accuracy, at best.

Finally, regarding binarity, the presence of even a wide binary during
the pre-main-sequence can prompt fast disk clearing, which could alter
a star's rotation period by halting disk-locking \citep{Meibom_2007}.
This mechanism might explain the abundance of fast rotators in
$\approx$120\,Myr open clusters \citep{Bouma_2021}.  A separate
concern with binaries is photometric blending of the rotation signal.
Because of these issues, our framework is only strictly applicable to
stars that are apparently single.
Section~\ref{subsec:binarityfilters} summarizes some of the
information that can be used to determine whether a given field star
meets this designation.  Appendix~\ref{app:includebinaries} discusses
the potential impact of ignoring binarity entirely.

\subsection{Future Directions}

\paragraph{The need for intermediate-age calibrators}
The region of Figure~\ref{fig:clusters} with the largest gap, near
1.8\,Gyr, has the largest systematic uncertainties in our model.
These uncertainties could be addressed by measuring rotation periods
in a cluster at this age.  Considering clusters from
\citet{CantatGaudin_2020} older than 1\,Gyr, within 1\,kpc, and with
more than 100 members yields eight objects.  Sorted near to far, they
are: Ruprecht-147, NGC-752, IC-4756, NGC-6991, NGC-2682, NGC-7762,
NGC-2423, and IC-4561.  The closest two have been studied by
\citet{Curtis_2020} and \citet{Agueros_2018}, though rotation periods
in NGC-752 ($1.34\pm0.06$\,Gyr, $d\sim440$\,pc) could be worth
revisiting using data from the Transiting Exoplanet Survey Satellite
and the	Zwicky Transient Facility.  IC-4756 and NGC-6991 could
similarly merit further study, though it would be wise to confirm
their ages before delving in a rotation period analysis.

\paragraph{Going older}
M67 (4\,Gyr) will likely be the next rung in the gyrochronology
ladder: the analyses by \citet{Barnes_2016} and \citet{Dungee_2022}
have nearly completed its rotation--color sequence.  As described in
Appendix~\ref{app:interp}, we used their data on M67 to calibrate the
rate of spin-down between 1 and 2.6\,Gyr.  This choice is connected
to a generic issue with interpolation-based methods: the systematic
uncertainty in the model increases near the boundaries of the
interpolation domain.  By this logic, incorporating the 4\,Gyr data in
the most reliable way would require an even older population of
stars.  Clusters such as NGC\,6791 (8\,Gyr; \citealt{Chaboyer_1999}),
or else a precise set of asteroseismic calibrators
\citep[\eg][]{vanSaders_2016} might be the most plausible paths toward
this goal, though the complicating effects of stellar evolution bear
consideration.

%\paragraph{Model improvements}
%Although our model fits the data, its assumed rotation period
%distribution at any given age and temperature
%(Equation~\ref{eq:residual}) is somewhat ad hoc.  For instance, we
%adopted a linear form for $g(t)$, the ``amplitude decay function''
%that forces the slow sequence to disappear by the age of Praesepe.
%Similarly, for $T_{\rm eff}^{\rm cut}(t)$, we simply asserted that
%this temperature at any given time was that of a star that just
%arrived at the ZAMS.  Other functional forms might provide better
%fits to the data, while perhaps providing greater explanatory power.
%A separate possible improvement concerns the choice of $\chi^2$ as
%the goodness of fit metric in Section~\ref{subsec:fitting}.
%Alternative techniques such as approximate Bayesian computation
%(ABC), combined with the use of for instance a two-sided
%Anderson-Darling test as a summary statistic
%\citep[\eg][]{Morris_2020} might be capable of capturing higher-order
%correlations between period and temperature otherwise omitted by the
%current approach.  Finally, it would certainly also be possible to
%improve the numerical efficiency of our implementation, for instance
%by adopting a back-end capable of symbolic differentiation, or by
%performing the integration over adaptive grids.

\paragraph{Precision age-dating of field stars}
The best way to demonstrate the reliability of a star's age is to
measure it using independent techniques.  One framework that we expect
to complement our own is the \texttt{BAFFLES} code
\citep{Stanford-Moore_2020}, which returns age posterior probabilities
based on a star's surface lithium content.  Other age-dating tools,
including activity (Ca HK, Ca IRT, x-ray, UV excess), isochrones, and
asteroseismology, can similarly be combined with our gyrochronal
posteriors to verify the accuracy of our rotation-based ages, and to
improve on their precision.

\citet{Angus_2019} presented an important step in this vein, through a
method that simultaneously fitted an isochronal and gyrochronal model
to determine a star's age.  Their statistical framework could
certainly encompass the model developed in this manuscript.  The main
advantages of our particular gyrochronology model however are {\it i)}
improved accuracy for stalling K-dwarfs, {\it ii)} improved accuracy
in treating the growth of the slow sequence and decay of the fast
sequence over the first gigayear, and {\it iii)} incorporation of the
astrophysical width of the slow sequence for FGK stars.  The main
disadvantage is that our model is not applicable beyond 2.6\,Gyr,
though we caution that this is because the calibration data are more
sparse in this regime, and so the ages have larger systematic
uncertainties.

\paragraph{Physics-based models}
A separate issue with our model is that it is empirical, and so it
does not yield physical understanding.  Physics-based gyrochronology
models have provided crucial insight into what gives the data in
Figure~\ref{fig:clusters} their structure.  The relevant physics
likely includes decoupling between the radiative core and convective
envelope \citep{Gallet_Bouvier_2013}, angular momentum transport to
recouple the core and envelope \citep{Gallet_Bouvier_2015,Spada_2020},
and spin-down rates that vary depending on whether the magnetic dynamo
is saturated \citep[\eg][]{Sills_2000,Matt_2015}.  At older ages,
additional physics may well be needed to explain the lethargic
spin-down of stars with Rossby numbers comparable to the Sun
\citep{Brown_2014,vanSaders_2016,David_2022}.  A separate issue that
also merits attention is the exact role of binarity on stellar
rotation.  Our filtering process
(Section~\ref{subsec:binarityfilters}) removed potential binaries
based on a gamut of tracers, because observations have shown that
rapid rotators are often binaries
\citep{Meibom_2007,Stauffer_2016_pleiades,Gillen_2020}.  The exact
properties of these binaries, for instance their separations and
masses, would help in clarifying the physical origin of this
correlation.  The issue of whether binarity leads to early disk
dispersal seems likely to be related, and also deserves attention
\citep{Cieza_2009}.

\acknowledgements
This work was supported by the 
Heising-Simons 51~Pegasi~b Fellowship (LGB)
and the Arthur R.~Adams SURF Fellowship (EKP).

\facilities{
  Gaia \citep{Gaia_DR3_2022},
  Kepler \citep{Borucki10},
  TESS \citep{ricker_transiting_2015},
  NGTS \citep{Wheatley_2018}
}

% \software{
%     astropy \citep{Astropy18},
%     matplotlib \citep{matplotlib},
%     numpy \citep{numpy},
%     scipy \citep{scipy},
% }

\clearpage

\bibliographystyle{aasjournal}
\bibliography{bibliography}

\appendix
\section{Interpolation Methods \& Literature Comparison}
\label{app:interp}

\begin{figure*}[!tpb]
	\begin{center}
		\leavevmode
		\subfloat{
			\includegraphics[width=0.33\textwidth]{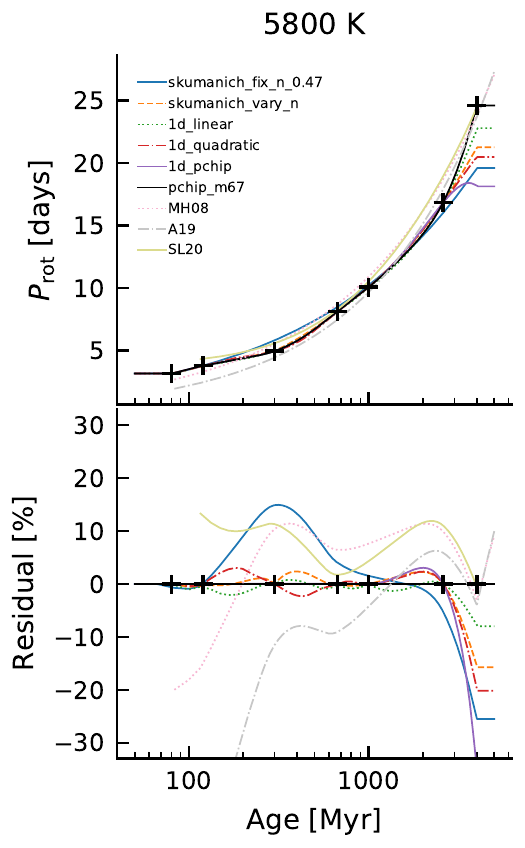}
			\includegraphics[width=0.33\textwidth]{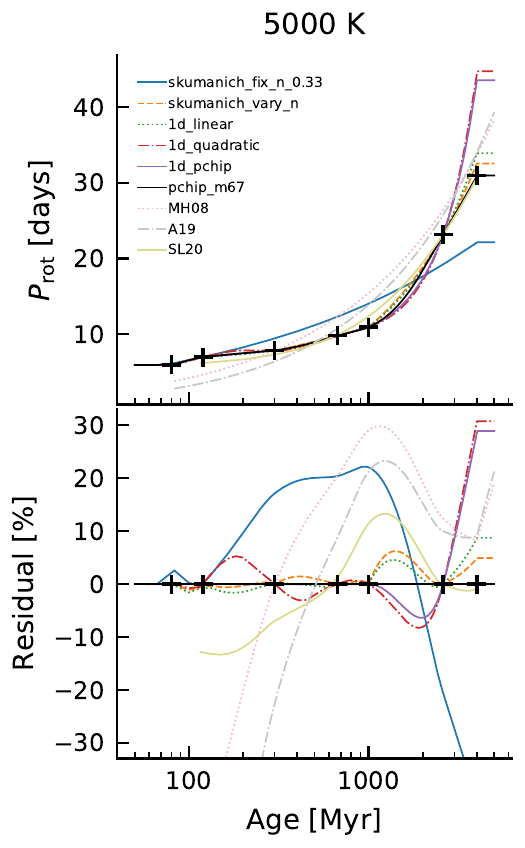}
			\includegraphics[width=0.33\textwidth]{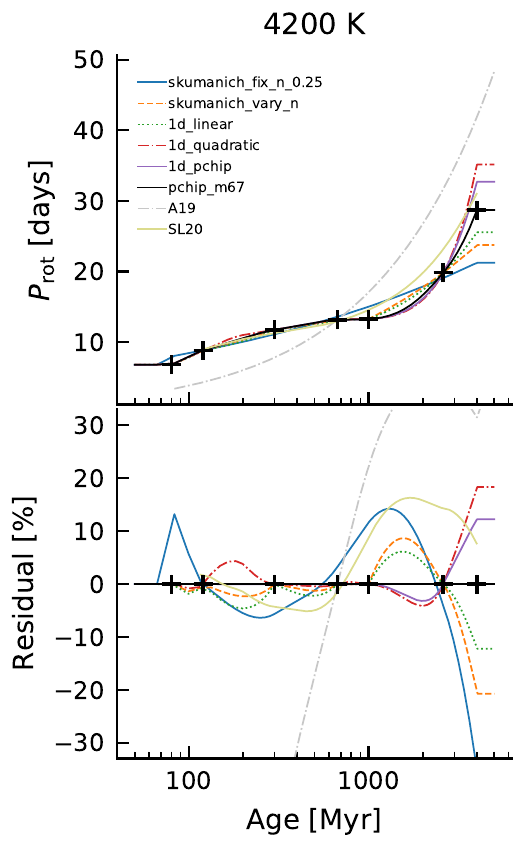}
		}
		
		\vspace{-0.4cm}
		\subfloat{
			\includegraphics[width=0.33\textwidth]{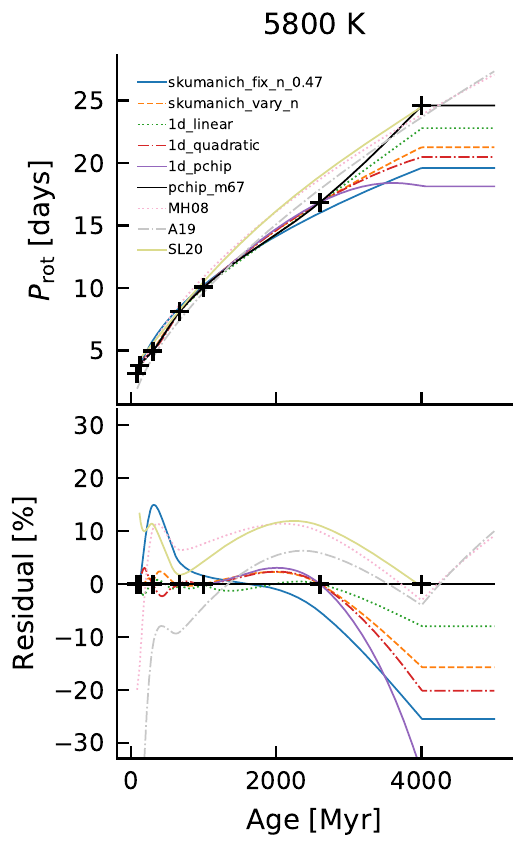}
			\includegraphics[width=0.33\textwidth]{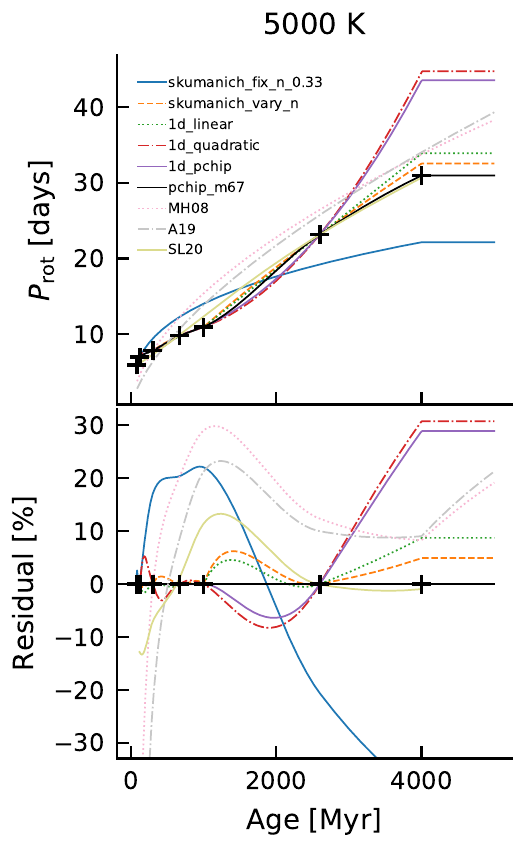}
			\includegraphics[width=0.33\textwidth]{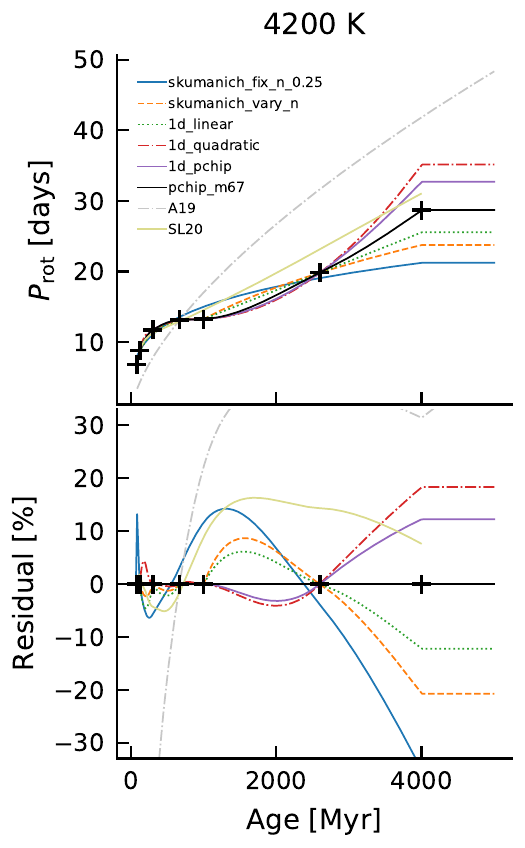}
		}
	\end{center}
	\vspace{-0.5cm}
	\caption{
    {\bf Different approaches for interpolating between reference
    clusters.}  $P_{\rm rot}$ denotes the rotation period of the star
    if it were evolving exactly along the slow sequence.  The top two
    and bottom two rows show identical data, but are scaled
    logarithmically and linearly in time.  The ``residual'' is defined
    versus the \texttt{pchip\_m67} interpolation method, calculated
    for each model $i$ as $(P_{\rm rot,i}-P_{\rm
    rot,pchip\_m67})/P_{\rm rot,i}$.  The ``+'' data points are
    evaluated from polynomial fits to the data in
    Figure~\ref{fig:clusters}. The fixed power laws
    (``\texttt{skumanich\_fix\_n\_0.XX}'') are extrapolated based on
    the rotation period at 120\,Myr.
    MH08: \citet{Mamajek_2008}.  A19: \citet{Angus_2019}.  SL20: \citet{Spada_2020}.
		\label{fig:prot_vs_age}
	}
\end{figure*}

How does the slow sequence evolve between each reference cluster?  In
other words, what is the functional form of $\mu_{\rm slow}(t, T_{\rm
eff})$, the rotation period of a star evolving exactly along the slow
sequence?  Figure~\ref{fig:prot_vs_age} summarizes a few possible
answers, evaluated at 5800\,K, 5000\,K, and 4200\,K.  Data from
\citet{Barnes_2016} and \citet{Dungee_2022} have been included as the
4\,Gyr M67 data points, to assess how well the interpolation methods
succeed at extrapolating beyond 2.6\,Gyr.
 As a qualitative note,
the stars of interest in this work (0.5--1.2\,$M_\odot$; 3800--6200\,K)
have temperature changes of $\lesssim$2.5\% between
80\,Myr and 2.6\,Gyr, according to the MIST grids \citep{Choi_2016}. 
Past $\approx$3\,Gyr, the most massive 1.2\,$M_\odot$ ($\approx 6200$\,K)
stars begin to turn onto the subgiant branch, and temperature becomes
a more ambiguous parameter when modeling stellar spin-down.

The simplest plausible model would be if the slow sequence's evolution
followed a power-law, with a flexible color or temperature calibration
similar to that suggested by many authors
\citep[\eg][]{Skumanich_1972,Noyes_1984,Barnes_2003}.  In this
approach, for every temperature, we would set $P_{\rm rot} \propto
t^n$, where canonically $n=1/2$.  We would then scale based on some
fiducial rotation period, say at 120\,Myr.
Figure~\ref{fig:prot_vs_age} shows how well this type of scaling
works, letting $n$ float in order to match the data as well as
possible.  For Sun-like stars ($\approx$$5800\,$K), this type of
scaling works surprisingly well, yielding agreement with the cluster
data at the $<$$15$\% level for $n=0.47$ out to 4\,Gyr.  The agreement
is significantly worse for lower mass stars, due to their stalled
spin-down at intermediate ages.

An alternative approach would be to directly interpolate between the
cluster sequences, ignoring our expectation for any kind of power-law
spin-down.  The resulting linear and quadratic interpolation cases are
shown as the dot and dot-dash lines in Figure~\ref{fig:prot_vs_age}.
While these approaches tautologically fit the data, they suffer from
sharp transitions in the spin-down rate at every reference cluster.
Quadratic interpolation is also not guaranteed to be monotonic, which
is probably a desirable property for a stellar spin-down.  A final
concern is that interpolating in this way is not guaranteed to be
predictive; leaving the M67 data out, extrapolating based on the
1--2.6\,Gyr data will generally over or under-estimate the rotation
periods in the 2.6--4\,Gyr interval.

An approach closer to interpolation that still incorporates a form of
power-law scaling is as follows.  For a point $(T_i,P_i)$ intermediate
between the loci of two clusters with ages $t_0$ and $t_1$ and
rotation periods $P_0$ and $P_1$ at the same temperature $T_i$, set
\begin{align}
  P_i &= {P_0} \left( \frac{t_i}{t_0} \right)^{n},
      \quad {\rm for}\quad
n = \frac{
    	\log\left( P_1 / P_0 \right)
}{
    	\log\left( t_1 / t_0 \right)
}.
\label{eq:n}
\end{align}
In other words, given the full set of reference loci $\{ \mu_0, \mu_1,
\ldots, \mu_k \}$, their ratios $\{ \mu_1/\mu_0, \ldots,
\mu_k/\mu_{k-1} \}$ can be used to define power-law scalings that are
accurate at a piecewise level.  While this tautologically fits the
data, there is a concern that for cool stars older than 1\,Gyr, it may
over-estimate the rotation periods.  This concern is primarily based
on the sharp transition visible in Figure~\ref{fig:prot_vs_age} in the
spin-down rate at 1\,Gyr for the 4200\,K case.

A final approach is based on PCHIP interpolation (Piecewise Cubic
Hermite Interpolating Polynomials; \citealt{Fitsch_1984}).  This
approach is monotonic, and continuous in the first derivatives at each
reference cluster.  While it is interpolation-based, and therefore not
predictive outside of its training bounds, we can include the M67 data
in order to define the most accurate possible slow sequence evolution
over the 1--2.6\,Gyr interval.  The results are shown with the black
line in Figure~\ref{fig:prot_vs_age} in the method labeled
``\texttt{pchip\_m67},'' which we adopt as our default.  This approach
leaves the slope of $P_{\rm rot}$ vs.~$t$ even less constrained in the
2.6--4\,Gyr interval, which is why we do not advocate using our model
for stars older than 2.6\,Gyr.

Finally, the models from \citet{Mamajek_2008} (MH08),
\citet{Angus_2019} (A19), and \citet{Spada_2020} (SL20) are also shown
in Figure~\ref{fig:prot_vs_age} for comparison.  The MH08 model is
defined over $0.5 < (B-V)_0 < 0.9$, or roughly 5050--6250\,K.  The
$T_{\rm eff}=5000\,{\rm K}$ case is therefore a mild
over-extrapolation, but we nonetheless show the result for
illustrative purposes.  Of the three cases, the \citet{Spada_2020}
model generally provides the best match to the data.

\section{What if we ignored binarity?}
\label{app:includebinaries}

In this work we argued that omitting binaries from gyrochronology
analyses is important due to the observational and astrophysical
biases that they can otherwise induce on rotation periods.  In
Section~\ref{subsec:binarityfilters}, we described the set of quality
filters that we used to expunge binaries from our calibration data,
to guarantee that we were considering only apparently single stars
with reliable rotation period measurements.  For generic field stars,
not all of these conditions are necessarily applicable.  For instance,
outliers in color--absolute magnitude diagrams might be challenging to
identify due to the lack of an immediately obvious reference sequence
(although the locus of the main-sequence is itself well-known, and
Gaia for instance can now be used to query local spatial volumes
around arbitrary field stars to construct well-defined reference
samples).

\begin{figure*}[!tb]
	\begin{center}
		\leavevmode
		\includegraphics[width=\textwidth]{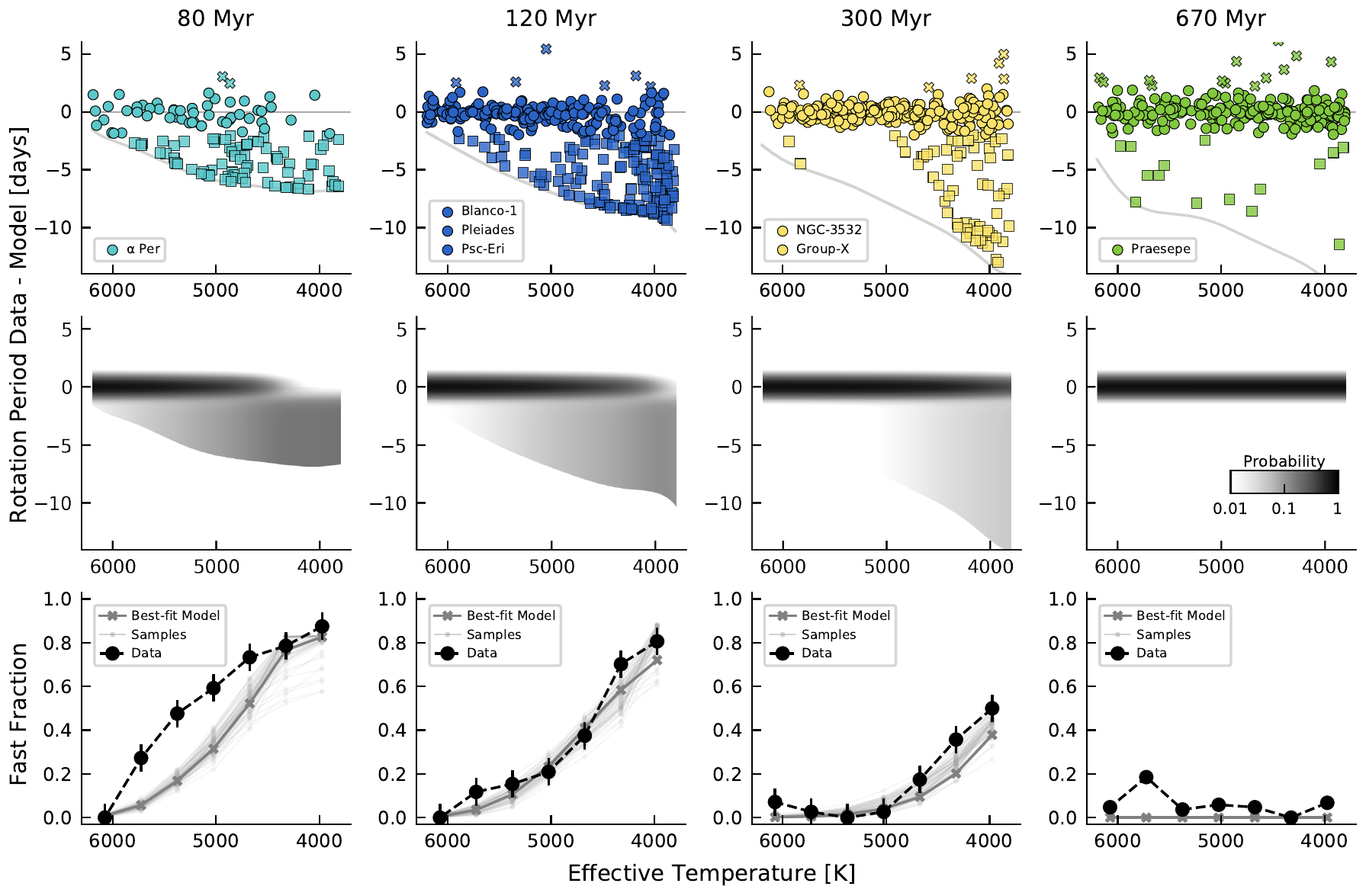}
	\end{center}
	\vspace{-0.5cm}
	\caption{
		{\bf What if we loosened the quality cuts?} 
    This plot is the same as Figure~\ref{fig:residual}, but
    systems that are known or suspected to be 
    visual, photometric, astrometric, and spectroscopic binaries
    are now displayed along with the single stars.
    The model is the same as in
    Figure~\ref{fig:residual}, as is the panel ordering.
		\label{fig:includebinaries}
	}
\end{figure*}

In general, we strongly recommend applying our method only to stars
that are thought to be single and on the main sequence.  For instance,
spectroscopic surface gravity estimates should be used, if available,
to expunge evolved stars since they are not in our calibration data.
Nonetheless, it is interesting to consider how well our method
translates for samples that are messier, and that have binarity rates
in line with field populations.  Figure~\ref{fig:includebinaries}
shows the result of dropping all of the quality cuts described in
Section~\ref{subsec:binarityfilters}, using the data included behind
Figure~\ref{fig:clusters}.

The first noticeable effect is that without any quality cuts, there
are more stars.  The star count in $\alpha$~Per jumps from 65 to 128;
in the 120\,Myr clusters from 196 to 364, in the 300\,Myr clusters
from 133 to 301; and in Praesepe from 100 to 250.  In addition,
without quality cuts, the width of the slow sequence increases.  The
mean residual width for the $t\geq120$\,Myr stars within 2~days of the
slow sequence is 0.72~days, a 40\% increase from $\sigma=0.51$\,days
observed in the cleaned sample.  This scatter term is proportional to
the statistical age uncertainty at late times, in the regime of very
precise rotation period and temperature measurements
\citep{Barnes_2007}.  This suggests that if one wished to apply our
gyrochronology model to a population with a mixture of single and
binary stars,  the model would need to be refit to account for the
wider intrinsic scatter in such a population.

Finally, we can ask to what degree the ratio between fast and slow
rotators changes when we omit all quality cuts.  The results are shown
in the bottom row of Figure~\ref{fig:includebinaries}, and compared
against the original best-fit model (trained on the cleaned data) from
Figure~\ref{fig:residual}.  While the visual agreement remains good at
$t\geq 120\,$Myr, the hot stars in the raw $\alpha$~Per sample have a
larger fast fraction than in the cleaned sample, and so the model
provides a worse match to those stars.  A second qualitatively
important difference is present in Praesepe: the raw data show around a
dozen rapid outliers, none of which are present in the cleaned dataset
(Figure~\ref{fig:residual}).  If any of these stars were single and
rapidly rotating, we might construe them as motivation to lengthen our
model's timescale for the decay of the fast sequence.  However, since
they are most likely binaries, and the Hyades similarly shows no
evidence for rapidly rotating single stars hotter than 3800\,K
\citep{Douglas_2019}.  The NGC-6811 data at 1\,Gyr similarly have no
reported rapid rotators \citep{Curtis_2019_ngc6811}.  We therefore
simply note that these outlying stars do exist at 0.7\,Gyr, and
that practitioners aiming to perform gyrochronology analyses on
populations of stars that include binaries should consider them.

\clearpage
\listofchanges

\end{document}